\documentclass[twocolumn,floatfix,preprintnumbers,nofootinbib,superscriptaddress,longbibliography]{revtex4-2}

\usepackage{ulem}
\usepackage{bm}
\usepackage{times}
\usepackage{amssymb,amsbsy,amsmath,amsfonts}
\usepackage{graphicx}
\usepackage{float}
\usepackage{color}
\usepackage{morefloats}
\usepackage{rotating}
\usepackage{srcltx}
\usepackage{slashed}
\usepackage{subfigure}
\usepackage{multirow}
\usepackage{verbatim}
\usepackage{hyperref}
\usepackage{tabularx}

\usepackage[outdir=./]{epstopdf}

\usepackage{graphicx}
\usepackage{amsmath}
\usepackage{amsfonts}
\usepackage{amssymb}
\usepackage{color}
\usepackage{multirow}
\usepackage{mathrsfs}
\usepackage{gensymb}

\usepackage{booktabs}  
\usepackage{threeparttable}  

\newcommand{\Slash}[1]{\ooalign{\hfil/\hfil\crcr$#1$}}

\usepackage{subfigure}

\newcommand\be{\begin{eqnarray}}
\newcommand\ee{\end{eqnarray}}

\usepackage{soul}

\begin{document}

\title{Role of $K^*_0(700)$ exchange in the $p \bar{p} \to \Lambda \bar{\Lambda}$ reaction}

\author{Hao-Nan Wang}
\affiliation{State Key Laboratory of Heavy Ion Science and Technology, Institute of Modern Physics, Chinese Academy of Sciences, Lanzhou 730000, China} 
\affiliation{School of Nuclear Sciences and Technology, University of Chinese Academy of Sciences, Beijing 101408, China}
\affiliation{Few-body Systems in Physics Laboratory, RIKEN Nishina Center, Wako, Saitama 351-0198, Japan}

\author{{Cheng Chen}}
\affiliation{State Key Laboratory of Heavy Ion Science and Technology, Institute of Modern Physics, Chinese Academy of Sciences, Lanzhou 730000, China} 
\affiliation{School of Nuclear Sciences and Technology, University of Chinese Academy of Sciences, Beijing 101408, China}

\author{Xing-Yi Ji}
\affiliation{School of Physics, Zhengzhou University, Zhengzhou 450001, China}
\affiliation{State Key Laboratory of Heavy Ion Science and Technology, Institute of Modern Physics, Chinese Academy of Sciences, Lanzhou 730000, China}

\author{De-Min Li}~\email{lidm@zzu.edu.cn}
\affiliation{School of Physics, Zhengzhou University, Zhengzhou 450001, China}

\author{Yue Ma}
\affiliation{Few-body Systems in Physics Laboratory, RIKEN Nishina Center, Wako, Saitama 351-0198, Japan}

\author{En Wang}~\email{wangen@zzu.edu.cn}
\affiliation{School of Physics, Zhengzhou University, Zhengzhou 450001, China}

\author{Ju-Jun Xie}~\email{xiejujun@impcas.ac.cn}
\affiliation{State Key Laboratory of Heavy Ion Science and Technology, Institute of Modern Physics, Chinese Academy of Sciences, Lanzhou 730000, China} 
\affiliation{School of Nuclear Sciences and Technology, University of Chinese Academy of Sciences, Beijing 101408, China}
\affiliation{Southern Center for Nuclear-Science Theory (SCNT), Institute of Modern Physics, Chinese Academy of Sciences, Huizhou 516000, China}

\begin{abstract}

Based on the effective Lagrangian approach, we investigate the $p \bar{p} \to \Lambda \bar{\Lambda}$ reaction. Within this framework, we provide a dynamical explanation by analyzing its total and differential cross sections, as well as the polarization of the produced $\Lambda$ hyperon. Incorporating the $t$-channel exchange of the scalar $K^*_0(700)$ and pseudoscalar $K$ mesons, complemented by an $s$-channel contribution from the vector excited state, we can reproduce the current experimental data fairly well in a wide energy region. Compared to the conventional $K$ and $K^*(892)$ mesons exchange, the $K^*_0(700)$ meson exchange plays a more essential role in simultaneously capturing the observed features of the total and differential cross sections. The introduction of the vector $s$-channel resonance improves the description of the spin observables. This work gives a perspective to inspect the role of $K^*_0(700)$ and serves as a test to search for the resonances in the reaction $p \bar{p} \to \Lambda \bar{\Lambda}$ at threshold.

\end{abstract}

\maketitle


\section{Introduction} \label{section:introduction}

The first baryon found to have the strange quark is the $\Lambda$ hyperon, and the $\Lambda \bar{\Lambda}$ hyperon-antihyperon production thus provides a good insight for understanding the intrinsic dynamics of hadrons including strangeness flavor. To date, a large number of experiments have measured the production of $\Lambda \bar{\Lambda}$. These experimental measurements played an essential role in establishing baryon-antibaryon interaction models and in elucidating the underlying quark-level interactions. The study of $\Lambda \bar{\Lambda}$ production in the $p \bar{p}$ scattering, which proceeds via strangeness-exchange mechanisms, has stimulated extensive investigations~\cite{Zhou:2022jwr}. During an earlier stage, many experiments successively measured the total cross sections of this reaction, providing data in an energy range of approximately 1~GeV above the $\Lambda \bar{\Lambda}$ threshold~\cite{Badier:1967zz,Oh:1973ny,Jayet:1978yq,Musgrave:1965zz,Jacobs:1977fq,Barnes:1987aw,Barnes:1989je,Barnes:1990bs,Barnes:1994bh}. Subsequently, based on the Low-energy Antiproton Ring (LEAR) of CERN, the production of $\Lambda \bar{\Lambda}$ has been studied by the LEAR PS185 experiment~\cite{Barnes:1989je,Barnes:1990bs,Barnes:1994bh,Barnes:1996si,Barnes:2000be}, and they gave the high statistical data of total and differential cross sections very close to the threshold of the $p \bar{p} \to \Lambda \bar{\Lambda}$ reaction in 2000~\cite{Barnes:2000be}. However, up to now, there has been no comprehensive theoretical analysis that can consistently account for all the experimental data from different experiments. Therefore, the primary objective of this work is to provide a unified description of these experimental measurements for the $p \bar{p} \to \Lambda \bar{\Lambda}$ reaction over a wide energy region, from which we can extract the energy dependence of the $p\bar{p}$ and $\Lambda\bar{\Lambda}$ interactions.

With the accumulation of experimental data of the $p \bar{p} \to \Lambda\bar{\Lambda}$ reaction, many theoretical models have been established, such as the $K$ and $K^*(892)$ strange meson-exchange model~\cite{Kohno:1987be,Mueller-Groeling:1990uxr,Haidenbauer:1991kt,Haidenbauer:1992wp,Haidenbauer:1992hv,Carbonell:1993dt,Shyam:2014dia}, partial wave analysis~\cite{Bugg:2004rj}, and one-gluon exchange model~\cite{Burkardt:1988pk}. As for the meson-exchange model, previous studies usually use the coupled channels with the effective potential to research the $\bar{K}$-nucleon interaction. This method also allows one to conveniently take into account the final-state interactions. In a meson-exchange description, $K$-exchange is the dominant long-range mechanism~\cite{Kohno:1987be}, and it was found that the $P$-wave contribution is important even very close to the reaction threshold. In Ref.~\cite{Ortega:2011zza}, a coupled channel calculation of the $p \bar{p} \to \Lambda \bar{\Lambda}$ reaction was performed in a constituent quark model which incorporates Goldstone boson exchanges between quarks and has been successfully applied to the $p \bar{p} \to p\bar{p}$ reaction. That work focuses on the study of depolarization and spin transfer observables in the $p \bar{p} \to \Lambda \bar{\Lambda}$ reaction at a $\bar{p}$ lab momentum $p_{\rm lab} = 1637$ MeV.

All these aforementioned model calculations can only describe the experimental data of the $p\bar{p} \to \Lambda \bar{\Lambda}$ reaction at low energies. But, the high energy data cannot be well described. Here, our work aims to investigate the total and differential scattering cross sections of this specific $p \bar{p} \to \Lambda \bar{\Lambda}$ process, in other words, we study it within a single-channel coupling framework. In this context, the effective Lagrangian approach turns out to be concise and efficient. Similarly, we also consider the $t$-channel $K$ and $K^*(892)$ exchanged mechanism. In addition, the scalar meson $K^*_0(700)$ (denoted as $\kappa$)~\footnote{In what follows, we will use $\kappa$ to denote $K^*_0(700)$ since the scalar meson $K^*_0(700)$ was also known as $\kappa$.} is also introduced. In fact, the $t$-channel scalar $\kappa$ meson exchange has been studied to interpret the hyperon-nucleon interaction~\cite{Rijken:1998yy,Rijken:2010zzb}, and the $\kappa$ meson has constituted an important component in the construction of the Nijmegen potential models~\cite{Rijken:1998yy,Rijken:2010zzb}. It was also pointed out, in Ref.~\cite{Ortega:2011zza}, the inclusion of the scalar strange $\kappa$ exchange at the quark level improves the description of the experimental data at $p_{\rm lab} = 1637$ MeV. Hence, our paper particularly pays attention to the contribution of the $\kappa$ meson to the process of $p \bar{p} \to \bar{\Lambda} \Lambda$ at the hadron level. The contributions from $t$-channel $K$ and $K^*(892)$ exchange are also discussed.

In addition to the $t$-channel strange meson exchange, we also consider the contributions from $s$-channel resonances. In fact, the near-threshold bound state of $\Lambda \bar{\Lambda}$ system can be used to explain the near-threshold enhancement of $e^+ e^- \to \Lambda \bar{\Lambda}$ reaction~\cite{BESIII:2017hyw,Li:2021lvs,Baldini:2007qg,Haidenbauer:2016won}. However, the dynamics near the $\Lambda \bar{\Lambda}$ threshold in the related reaction $p \bar{p} \to \Lambda \bar{\Lambda}$ may be distinct. The recent PS185 experiment has given a precise measurement of $p \bar{p} \to \Lambda \bar{\Lambda}$ reaction in the kinematic region from the reaction threshold up to an excess energy of about 6~MeV~\cite{Barnes:2000be}. They found that the total cross section varied smoothly near the threshold, indicating no observable resonant structures. A detailed study of the shape of the total cross section should reveal a structure if such an intermediate state is an important part of the reaction dynamics of the $p \bar{p} \to \Lambda \bar{\Lambda}$ reaction. Here, we conduct a comprehensive analysis of intermediate resonance states based on experimental data of both total and differential cross sections. This investigation is particularly significant as, despite their theoretically suppressed contributions predicted by the Okubo-Zweig-Izuka rule (contribution arising from non-planar-quark diagram), these resonances may still play a non-negligible role in some energy regions. Thus, in the present paper, we investigate the $p\bar{p}\to \Lambda \bar\Lambda$ reaction by introducing $t$-channel $K$, $K^*(892)$, and $\kappa$ exchange, as well as an $s$-channel resonance within the effective Lagrangian approach. Our focus is on describing the total and differential cross-section data over a broad energy range. 

On the other hand, the $p \bar{p} \to \Lambda \bar{\Lambda}$ reaction provides an ideal laboratory for investigating spin dynamics, primarily due to the self-analyzing nature of the parity-violating weak decays of $\Lambda$ and $\bar{\Lambda}$ hyperons~\cite{Barnes:1990bs}. Additionally, Ref.~\cite{Alberg:1995zp} points out that measurements of the depolarization in the $p \bar{p} \to \Lambda \bar{\Lambda}$ process can serve as an interesting test of the dynamics responsible for the "proton spin puzzle". Within the meson-exchange method, numerous works have investigated the associated spin observables~\cite{LaFrance:1990kb,Timmermans:1992fu,Ortega:2011zza,Zhou:2013ioa} in $p\bar{p}$ reactions. For the $p \bar{p} \to \Lambda \bar{\Lambda}$ reaction, a full coupled-channel approach strictly satisfying unitarity was used to investigate $\Lambda$ polarization in Ref.~\cite{Haidenbauer:1991kt}, while the distorted-wave Born approximation with optical potentials was employed to account for the spin dynamics~\cite{LaFrance:1990kb}. The contribution of $K^*(892)$ exchange was also examined in Ref.~\cite{LaFrance:1990kb}. The authors demonstrated that comparing the fitting parameters extracted from different $KN\Lambda$ and $K^*N\Lambda$ couplings schemes serves to verify the internal consistency of the model. To reproduce the same experimental measurements, a reasonable balance between production and absorption in the combined $K$ and $K^*(892)$ sectors has to be respected. In the present work, we adopt a phenomenological effective Lagrangian approach with spin density matrix~\cite{Fano:1983zz,Tabakin:1991ct}, which is obtained from the scattering amplitude to investigate $\Lambda$ polarization. Note that the polarization observables arise directly from the interference of scattering amplitudes, making them a sensitive probe of the relative phases and reaction mechanisms. Thus, relative phases between different contributions are included.

This article is organized as follows. In Section~\ref{sec:formalism}, we provide a detailed description of our theoretical framework, and outline the effective Lagrangian model used to calculate the cross sections of the $p \bar{p} \to \Lambda \bar{\Lambda}$ reaction. Section~\ref{sec:results} presents the fitting of experimental data based on this theoretical framework, accompanied by a discussion of the results. Finally, Section~\ref{sec:summary} offers a short summary.


\section{Theoretical formalism} \label{sec:formalism}

\begin{figure}[htbp]
\centering
\subfigure[]{\includegraphics[scale=0.43]{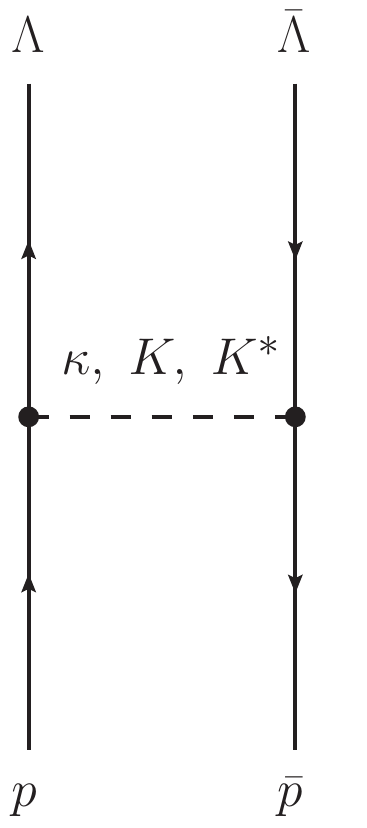}
\label{fig:kappa}} \hspace{1.5cm}
\subfigure[]{\includegraphics[scale=0.43]{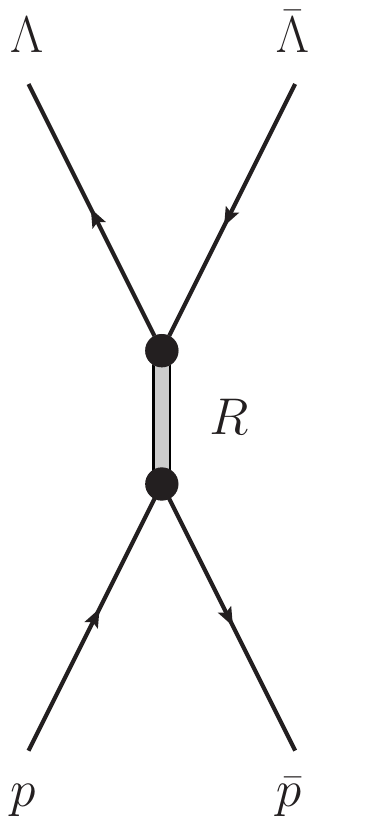}
\label{fig:HX}}
\caption{Feynman diagrams for the $p \bar{p} \to \Lambda \bar{\Lambda}$ reaction. (a) $t$-channel with scalar $\kappa$, pseudoscalar $K$, and vector $K^*$; (b) $s$-channel with resonance $R$.}
\label{fig:feynman}
\end{figure}

As mentioned in the introduction, we will employ an effective Lagrangian method to investigate the $p \bar{p} \to \Lambda \bar{\Lambda}$ reaction. In the following analysis, we consider a combined contribution from both the $t$-channel and $s$-channel processes, as shown in FIG.~\ref{fig:feynman}. For the $t$-channel exchange, we sequentially study the role of scalar $\kappa$, pseudoscalar $K$, and vector $K^*(892)$ ($\equiv K^*$), in order to compare their respective effects and differences. There is no signal for $s$-channel resonance contribution to the $p \bar{p} \to \Lambda \bar{\Lambda}$ reaction near the threshold~\cite{Barnes:2000be}, while there is an evidence for a resonance contribution in the high energy region. Here, we will also study the important role played by the excited meson states with masses around 2.2 to 3.0 GeV, with the aim of describing the available experimental data of the $p \bar{p} \to \Lambda \bar{\Lambda}$ reaction at the high energy region. The $R$ in FIG.~\ref{fig:feynman} stands for the $s$-channel resonance. Consequently, we define three representative scenarios for comparison: Set I, including $\kappa$, $K$, and $R$; Set II, including $\kappa$, $K^*$, and $R$; Set III, including $K$, $K^*$, and $R$.. For doing this, we have minimized the free model parameters in order to have the best reproduction of the available experimental data.

Since the information about the meson states around $2.2$ to $3.0$ GeV is scarce~\cite{ParticleDataGroup:2024cfk}, it is necessary to rely on theoretical calculations. For example, the spectrum of excited $\omega$ and $\phi$ states was investigated in Ref.~\cite{Wang:2021abg} using the modified Godfrey-Isgur model. By performing phenomenological analysis, the two-body $\Lambda \bar{\Lambda}$ strong decays of the excited $\phi$ states were studied in Ref.~\cite{Bai:2023dhc}. Based on the above theoretical calculations, we take one excited vector state~\footnote{The $p\bar{p}$ system annihilates with the angular momentum $l=0$ can also have quantum numbers $J^{PC} = 0^{-+}$, which means the intermediate state could also be an excited pseudoscalar meson.} into account in this work. Its mass and width are determined to reproduce the available experimental data for the $p \bar{p} \to \Lambda \bar{\Lambda}$ reaction.

To compute the contributions of $t$-channel $\kappa$, $K$, and $K^*$ exchange, and $s$-channel $R$ resonance, we use the interaction Lagrangian densities of Refs.~\cite{Kim:2011rm,Xie:2011me,Xie:2010yk,Xie:2013wfa,Wang:2016fhj,Wang:2017hug,Xie:2013mua,Xie:2014kja}:
\be 
    && \mathcal{L}_{\kappa} = - g_{\kappa p \Lambda} \bar{\psi}_{\Lambda} \psi_p\phi_{\kappa} + {\rm H.c.} \label{eq:kappa}, \\
    && \mathcal{L}_{K} = - g_{K p \Lambda}\bar{\psi}_{\Lambda} \gamma_5\psi_p\phi_K + {\rm H.c.}, \\
    && \mathcal{L}_{K^*} = g_{K^* p \Lambda} \bar{\psi}_{{\Lambda}} \gamma_{\mu} \psi_p \phi^{\mu}_{K^*} \notag\\
     && ~~~~~~~~~~~~ + \frac{f}{4m}\bar{\psi}_{{\Lambda}} \sigma_{\mu\nu} \psi_p F^{\mu\nu}_{K^*} + {\rm H.c.}, \\
    && {\mathcal{L}_{R p\bar{p}} = g_{R p\bar{p}} \phi_R^{\mu}\bar{\psi}_{\bar{p}} \gamma_{\mu} \psi_p } \notag\\
    && { ~~~~~~~~~~~~~~ + \frac{f_{R p\bar{p}}}{2(m_p + m_{\Lambda})}\bar{\psi}_{\bar{p}} \sigma_{\mu\nu} \psi_p F^{\mu\nu}_{R} + {\rm H.c.},  \label{eq:Xppnew}} \\    
   && {\mathcal{L}_{R \Lambda \bar{\Lambda}} = g_{R \Lambda \bar{\Lambda}} \phi_R^{\mu}\bar{\psi}_{\Lambda} \gamma_{\mu} \psi_{\bar{\Lambda}} } \notag\\
    && { ~~~~~~~~~~~~~~ + \frac{f_{R \Lambda \bar{\Lambda}}}{2(m_p + m_{\Lambda})}\bar{\psi}_{{\Lambda}} \sigma_{\mu\nu} \psi_{\bar{\Lambda}} F^{\mu\nu}_{R} + {\rm H.c.},
    \label{eq:XLLnew}}
\ee 
where the $\mathcal{L}_{\kappa}$, $\mathcal{L}_{K}$, and $\mathcal{L}_{K^*}$ describe the interactions of $p \bar{p} \to \Lambda \bar{\Lambda}$ in $t$-channel with scalar meson $\kappa$, pseudoscalar meson $K$, and vector meson $K^*$, respectively. $\phi^{\mu}_{K^*}$ and $\phi^{\mu}_R$ indicate the polarization vectors for the exchanged $K^*$ meson and $s$-channel $R$ resonance, respectively. $g_{\kappa p \Lambda}$ is the coupling constant of vertex $\kappa p \Lambda$, $g_{K p \Lambda}$ is corresponding to $K p \Lambda$ interaction, while $g_{K^* p \Lambda}$, and $f$ are the coupling constants depicting $K^* p \Lambda$ interaction. The corresponding Feynman diagram is shown as FIG.~\ref{fig:kappa}. Similar to the $K^*$ situation, as for resonance $R$, $g_{R p\bar{p} / R \Lambda \bar{\Lambda}}$ and $f_{R p\bar{p} / R \Lambda \bar{\Lambda}}$ are the coupling constants of the vector term and tensor term, respectively. Based on the investigations of $NK$ interactions in Ref.~\cite{Mueller-Groeling:1990uxr}, we take $g_{K p \Lambda} = 13.98$.~\footnote{Note that in our fitting procedure, we treat the coupling constant $g_{Kp\Lambda}$ as a free fitting parameter, and its fitted value consistently lies close to $14$. Given its good agreement with the value of $13.98$, we therefore decided to fix this coupling at $13.98$. This treatment not only maintains theoretical consistency but also reduces the number of free parameters, thereby effectively suppressing the possibility of over-fitting.}  

As for the $s$-channel, $\mathcal{L}_{R p\bar{p}}$ in Eq.~(\ref{eq:Xppnew}) and $\mathcal{L}_{R \Lambda \bar{\Lambda}}$ in Eq.~(\ref{eq:XLLnew}) describe the interactions between the vector resonance $R$ and $p \bar{p}$ and $\Lambda \bar{\Lambda}$, respectively. And the resonance $R$ in $s$-channel, with quantum numbers $J^{PC} = 1^{--}$, is shown as FIG.~\ref{fig:HX}. { In the study of $NK$ interaction results~\cite{Mueller-Groeling:1990uxr}, all of the coupling constants are real. However, within our theoretical framework, relying solely on the interference between different particles is insufficient to describe the spin polarization. Consequently, the interference between the vector and tensor couplings of the $K^* / R$ vector mesons becomes particularly important. Finally, for the sake of simplicity and effectiveness, we treat the tensor couplings $f$, $f_{R p\bar{p} / R \Lambda \bar{\Lambda}}$ as the complex parameters. For the $s$-channel resonance, since the squared momentum transfer $q^2 > 0$ lies in the time-like region, the coupling constants of the resonance $R$ to the baryon-antibaryon pair can in principle be complex. Regarding the $t$-channel $K^*$ exchange, its vector and tensor coupling constants are typically treated as real numbers in conventional studies~\cite{Mueller-Groeling:1990uxr,Nagels:1979xh,Kohno:1987uj,LaFrance:1990kb}. However, within our phenomenological framework, we intentionally introduce a complex tensorial coupling for the $K^*$ to introduce a relative phase between its vector and tensor coupling terms. This is aimed at providing maximum flexibility in describing the polarization observables of $\Lambda$ hyperon, as will be demonstrated in the subsequent discussions. Furthermore, the effects of absorption and unitarity are incorporated by introducing phenomenological imaginary tensorial couplings for both the $K^*$ and the $R$ resonance. All these coupling parameters will be determined by fitting them to the experimental data.

In our scattering amplitude calculation of Feynman diagrams shown in Fig.~\ref{fig:feynman}, we also need the propagators of scalar $\kappa$, pseudoscalar $K$, vector $K^*$, and resonance $R$. Then their propagators are given by
\be 
    && G_{\kappa / K} = \frac{i}{q^2_{\kappa / K} - m_{\kappa / K}^2}, \\
    && G_{K^*}^{\mu\nu} = i\frac{g^{\mu\nu} - q_{K^*}^{\mu} q_{K^*}^{\nu} / m^2_{K^*}}{q_{K^*}^2 - m_{K^*}^2 }, \\
    && G_{R}^{\mu\nu} = i\frac{g^{\mu\nu} - q_{R}^{\mu} q_{R}^{\nu} / m^2_{R}}{q_{R}^2 - m_{R}^2 + i m_{R} \Gamma_{R}},
\ee 
where $q_{\kappa / K/K^*}$ and $m_{\kappa / K/K^*}$ are the momentum and mass of $t$-channel exchanged $\kappa$ ($K$ or $K^*$), respectively. In this work we take $m_{\kappa} = 838$ MeV, $m_K = 495.6$ MeV, and $m_{K^*} = 893.6$ MeV as in the Review of Particle Physics (RPP)~\cite{ParticleDataGroup:2024cfk}. Similarly, $q_R$, $m_R$, and $\Gamma_R$ represent the momentum, mass, and width of the $s$-channel resonance, respectively, where $m_R$ and $\Gamma_R$ are model parameters and they will be obtained by fitting to the available experimental data.

Except for the propagators of the exchanged particles, it is also necessary to introduce the form factor to modify the impact of off-shell behavior of the exchanged particles~\cite{Liu:1995st,Xie:2007qt,Xie:2015zga,Zhao:2019syt}. There is no unique theoretical way to introduce the form factors~\cite{Xie:2010yk}. We adopt here the scheme used in the previous works which can be regulated as follows~\cite{Machleidt:1987hj,Xie:2014zga}:
\be 
    F_{\kappa / K / K^*}(q_{\kappa / K / K^*}) = \left( \frac{\lambda_{\kappa / K / K^*}^2}{\lambda_{\kappa / K / K^*}^2 - q_{\kappa / K / K^*}^2} \right)^n,
\ee 
for the $t$-channel with the $\kappa$, $K$, $K^*$ cases, we take $n=1$ for $\kappa$ and $K$, and $n=2$ for the vector $K^*$. And
\be
    F_{R}(s) = \frac{\lambda_{R}^4}{\lambda_R^4 + (s - m_R^2)^2},
\ee 
for the $s$-channel with the vector meson $R$ case~\cite{Xie:2005sb,Xie:2007vs,Wang:2014jxb,Dai:2025hvo}. And $s=q^2_R$ is the square of the center-of-mass energy of the $p \bar{p}$ system. We consider different parmetrizations of the form factors with the cut-off parameters $\lambda_{\kappa / K /K^*}$ and $\lambda_R$. Note that the values of the cutoff parameters can be directly related to the hadron size. Nevertheless, the hadron size is still an open issue in hadron physics. Therefor, we phenomenologically tune these cutoff parameters to reproduce available experimental data. In the present work, we determine the values of these parameters by fitting to the differential and total cross sections, as well as the spin observables of the $p\bar{p} \to \Lambda \bar{\Lambda}$ reaction.

With the effective interaction Lagrangian and form factors given above, we can easily construct the invariant scattering amplitude for Set I and Set II
\be 
  && {\mathcal{M} = \mathcal{A}^{\kappa} F_{\kappa}(q_{\kappa}) + \mathcal{A}^{ K / K^* } F_{ K / K^* }(q_{ K / K^* }) e^{i \varphi_1} } \notag\\
    && ~~~~~~~~~ { + \mathcal{A}^{ R } F_{ R }(q_{ R }) e^{i \varphi_2},}  \label{eq:isamplitude}
\ee
{where $\varphi_{1}$ and $\varphi_{2}$ represent the relative phases between the scattering amplitudes of $K$ (or $K^*$) and $\kappa$, and between those of resonance $R$ and $\kappa$, respectively. Similarly, the amplitude for Set III can be written as ${\mathcal{M} = \mathcal{A}^{K} F_{K}(q_{K}) + \mathcal{A}^{ K^* } F_{ K^* }(q_{ K^* }) e^{i \varphi_1} } + { \mathcal{A}^{ R } F_{ R }(q_{ R }) e^{i \varphi_2}}$, currently $\varphi_{1}$ and $\varphi_{2}$ represent the relative phase between the scattering amplitudes of $K$ and $K^*$, and between those of resonance $R$ and $K$, respectively. In Set I, we consider the contributions from $\kappa$, $K$, and resonance $R$, whereas in Set II we replace $K$ with $K^*$. By artificially turning off the $K$ exchange and leaving only the $K^*$ exchange in Set II, we aimed to isolate and examine the solitary effect of $K^*$. Moreover, we can investigate the role of $\kappa$ by turn off it in Set III.

The reduced amplitudes $\mathcal{A}^{\kappa / K / K^* / R}$ shown in Eq.~\eqref{eq:isamplitude} depend on the spin variables of the initial proton and antiproton and the final $\Lambda$ and $\bar{\Lambda}$, which are given by as follows:}
\be 
   &&  \mathcal{A}^{\kappa}_{s_p,s_{\bar{p}},s_{\Lambda},s_{\bar{\Lambda}}} =  \frac{ i g^2_{\kappa p \Lambda}}{q^2_{\kappa} - m_{\kappa}^2} \times \notag \\
     && \! \! \! \! \! \! \! \! \Big[ \bar{u}_{\Lambda}(p_{\Lambda},s_\Lambda) u_p(p_p,s_p)  \bar{v}_{\bar{p}}(p_{\bar{p}},s_{\bar{p}}) v_{\bar{\Lambda}}(p_{\bar{\Lambda}},s_{\bar{\Lambda}}) \Big], \\
&& \mathcal{A}^{K}_{s_p,s_{\bar{p}},s_{\Lambda},s_{\bar{\Lambda}}}=  \frac{ i g^2_{K p \Lambda}}{q^2_{K} - m_{K}^2} \times \notag \\
     && \! \! \! \! \! \! \! \! \! \! \! \! \! \! \! \! \! \! \! \!  \Big[ \bar{u}_{\Lambda}(p_{\Lambda},s_\Lambda) \gamma_5 u_p(p_p,s_p)  \bar{v}_{\bar{p}}(p_{\bar{p}},s_{\bar{p}}) \gamma_5 v_{\bar{\Lambda}}(p_{\bar{\Lambda}},s_{\bar{\Lambda}}) \Big], \\
&& \mathcal{A}^{K^*}_{s_p,s_{\bar{p}},s_{\Lambda},s_{\bar{\Lambda}}} =  \frac{i}{q_{K^*}^2 - m_{K^*}^2} \notag\\
     && \times \Big[ g_{K^* p \Lambda} \bar{u}_{{\Lambda}}(p_{\Lambda},s_{\Lambda}) \gamma_{\mu} u_p(p_p, s_p) \epsilon^{*\mu}_{K^*} \notag\\
     && + \frac{f}{4m}\bar{u}_{\Lambda}(p_{\Lambda}, s_{\Lambda}) \sigma_{\mu\nu} u_p(p_p, s_{p}) F^{\mu\nu}_{K^*} \Big] \notag\\
     && \times \Big[ g_{K^* p \Lambda} \bar{v}_{\bar{p}}(p_{\bar{p}}, s_{\bar{p}}) \gamma_{\mu'} v_{\bar{\Lambda}}(p_{\bar{\Lambda}}, s_{\bar{\Lambda}}) \epsilon^{\mu}_{K^*} \notag\\
     && + \frac{f}{4m}\bar{v}_{\bar{p}}(p_{\bar{p}}, s_{\bar{p}}) \sigma_{\mu'\nu'} v_{\bar{\Lambda}}(p_{\bar{\Lambda}}, s_{\bar{\Lambda}}) F^{\mu'\nu'}_{K^*} \Big], \\
    && \mathcal{A}^{R}_{s_p,s_{\bar{p}},s_{\Lambda},s_{\bar{\Lambda}}}   =   \frac{ i }{q_{R}^2 - m_{R}^2 +  i m_{R} \Gamma_{R}}  \notag\\
     && \times \Big[ g_{ R \Lambda \bar{\Lambda} } \bar{u}_{{\Lambda}}(p_{\Lambda},s_{\Lambda}) \gamma_{\mu} v_{\bar{\Lambda}}(p_{\bar{\Lambda}}, s_{\bar{\Lambda}}) \epsilon^{*\mu}_{R}  \notag\\
     &&  + \frac{f_{R \Lambda \bar{\Lambda}}}{4m}\bar{u}_{\Lambda}(p_{\Lambda}, s_{\Lambda}) \sigma_{\mu\nu} v_{\bar{\Lambda}}(p_{\bar{\Lambda}}, p_{\bar{\Lambda}}) F^{\mu\nu}_{R} \Big] \notag\\
     && \times \Big[ g_{R p \bar{p}} \bar{v}_{\bar{p}}(p_{\bar{p}}, s_{\bar{p}}) \gamma_{\mu'} u_{p}(p_{p}, s_{p}) \epsilon^{\mu'}_{R}  \notag\\
     &&  + \frac{f_{R p \bar{p}}}{4m}\bar{v}_{\bar{p}}(p_{\bar{p}}, s_{\bar{p}}) \sigma_{\mu'\nu'} u_{p}(p_{p}, s_{p}) F^{\mu'\nu'}_{R} \Big], 
\ee
where $p_i$ and $s_i$ ($i=p$, $\bar{p}$, $\Lambda$, $\bar{\Lambda}$) stand for the momentum and spin of the corresponding particle, respectively, and $q_{\kappa / K/K^*} = p_p - p_{\Lambda}$.

After establishing the complete scattering amplitudes, the calculation of the invariant amplitude square $|{\cal M}|^2$ and the differential cross section of $p \bar{p} \to \Lambda \bar{\Lambda}$ reaction is straightforward~\cite{ParticleDataGroup:2024cfk}, 
\be 
   \frac{{\rm d}\sigma}{{\rm d}\cos{\theta}} = \frac{|\boldsymbol{p}_{\Lambda}|}{32\pi s |\boldsymbol{p}_p| }\frac{1}{4}\sum_{s_p,s_{\bar{p}},s_\Lambda,s_{\bar{\Lambda}}}|\mathcal{M}|^2,
\ee 
where $\boldsymbol{p}_{\Lambda}$ and $\boldsymbol{p}_p$ represent the three-momentum of proton and hyperon in the mass-center system, respectively, and $\theta$ denotes the angle between the above two momenta. $|\boldsymbol{p}_{\Lambda}|$ and $|\boldsymbol{p}_p|$ are given by
\be 
 |\boldsymbol{p}_{\Lambda}| = \frac{\sqrt{s-4m^2_{\Lambda}}}{2}, ~~~~ |\boldsymbol{p}_p| = \frac{\sqrt{s-4m^2_p}}{2} ,
\ee
with $m_p = 938.272$ MeV and $m_{\Lambda} = 1115.683$ MeV.

\begin{figure}[htbp]
\centering
\includegraphics[scale=0.6]{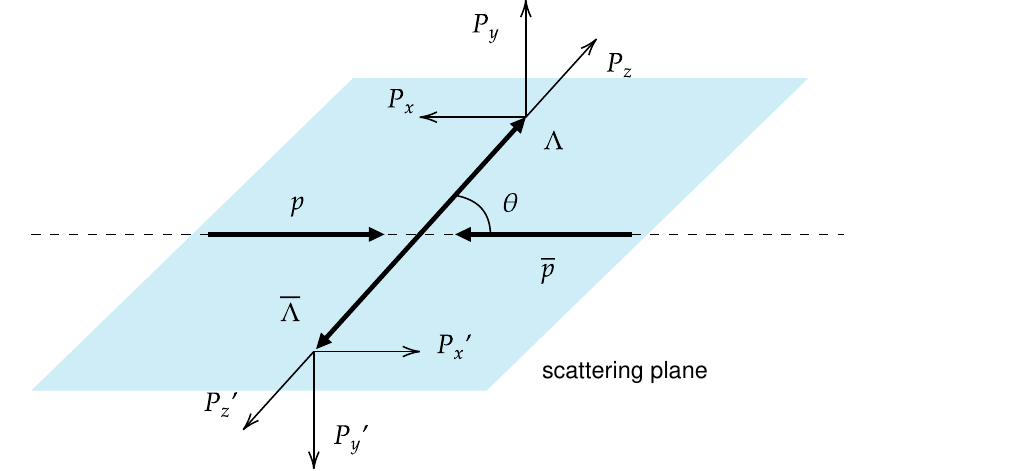}
\caption{{Definition of the scattering plane in the $p \bar{p}$ and $\Lambda \bar{\Lambda}$ mass-center system. $P_y$ and $P'_y$ are the polarizations along their $y$-axis, respectively.}}
\label{fig:scattering_plane}
\end{figure}

\begin{table*}[!htbp]
\renewcommand\arraystretch{1.5}
\centering
		\caption{ {Fitted parameters in this work. There are three fitting scenarios: Set I ($\kappa$, $K$, and $R$), Set II ($\kappa$, $K^*$, and $R$), Set III ($K$, $K^*$, and $R$).} }
		\label{table:fit_arameters} 
		\begin{tabular}{ccccc}   
			\toprule[1.5pt]
			 \multicolumn{2}{c}{~~~~~~parameter~~~~~~} & ~~~~~~Set I~~~~~~ & ~~~~~~Set II~~~~~~ & ~~~~~~Set III~~~~~~ \\
			 \hline
			 \multirow{2}{*}{$\kappa$} & $g_{\kappa p \Lambda}$ & $7.5 \pm 0.3$ & $4.6 \pm 0.1$ & -- \\  \multirow{2}{*}{ } & $\lambda_{\kappa}$[MeV] & $121 \pm 4$ & $243 \pm 4$ & -- \\
    \hline
        \multirow{1}{*}{$K$} & $\lambda_{K}$~[MeV] & $181 \pm 5$ & -- & $206 \pm 3$ \\
        \hline
        \multirow{3}{*}{$K^*$} & $g_{K^* p \Lambda}$ & -- & $1.2 \pm 0.2$ & $1.1 \pm 0.1$  \\
        \multirow{3}{*}{} & $f$ & -- & $(-9.9 \pm 0.7) + i(3.2 \pm 1.5)$ & $(-2.7 \pm 0.2) + i(2.1 \pm 0.5)$ \\
        \multirow{3}{*}{} & $\lambda_{K^*}$~[MeV] & -- & $485 \pm 8$ & $1009 \pm 90$ \\
        \hline
			 \multicolumn{2}{c}{$m_{R}$~[MeV]} & $2130 \pm 5$ & $1986 \pm 25$ & $2138 \pm 2$ \\
			 \multicolumn{2}{c}{$\Gamma_{R}$~[MeV]} & $292 \pm 18$ & $26 \pm 6$ & $1600 \pm 70$ \\
             \multicolumn{2}{c}{$\lambda_{R}$~[MeV]} & $337 \pm 9$ & $638 \pm 9$ & $281 \pm 1$ \\
             \hline
             \multicolumn{2}{c}{$g_{R p \bar{p}}$}   & $236.7 \pm 2.1$ & $72.6 \pm 1.4$ & $3833 \pm 22$ \\
             \multicolumn{2}{c}{$f_{R p \bar{p}}$}   & $(-175.4 \pm 2.6) - i(24.0 \pm 4.0)$ & $(-56.6 \pm 2.0) + i(27.0 \pm 1.7)$ & $(-3140 \pm 21) - i(129 \pm 20)$ \\
             \hline
             \multicolumn{2}{c}{$g_{R \Lambda \bar{\Lambda}}$}   & $14.3 \pm 0.01$ & $8.9 \pm 0.01$ & $9.2 \pm 0.01$ \\
             \multicolumn{2}{c}{$f_{R \Lambda \bar{\Lambda}}$}   & $(-14.4 \pm 0.01) - i(0.3 \pm 0.03)$ & $(-8.9 \pm 0.01) + i(0.2 \pm 0.01)$ & $(-9.2 \pm 0.01) - i(0.1 \pm 0.01)$ \\
             \hline
             \multicolumn{2}{c}{$\varphi_1 ~(\degree)$}   & $62.6 \pm 1.4$ & $-1.0 \pm 1.1$ & $117.5 \pm 2.3$ \\
             \multicolumn{2}{c}{$\varphi_2 ~(\degree)$}   & $179.2 \pm 1.8$ & $60.6 \pm 0.01$ & $17.2 \pm 0.7$ \\
             \hline
             \multicolumn{2}{c}{$\chi^2/{\rm d.o.f}$} & $2.2$ & $2.7$ & $2.2$ \\
			\toprule[1.5pt]
		\end{tabular}    
\end{table*}

{To calculate the spin observables, we evaluate the matrix elements of the invariant scattering amplitudes using explicit Dirac spinors. Due to parity conservation in the strong interaction, the polarization of the outgoing hyperon for unpolarized initial states is non-vanishing only along the direction normal to the scattering plane (i.e., $y$-axis, with positive direction defined by $\textbf{e}_y = {\boldsymbol{p}}_{p} \times {\boldsymbol{p}}_{\Lambda}/|{\boldsymbol{p}}_{p} \times {\boldsymbol{p}}_{\Lambda}| $), as shown in FIG.~\ref{fig:scattering_plane}. Consequently, we focus on the spin projections along this axis. In this $p \bar{p} \to \Lambda \bar{\Lambda}$ reaction, the Dirac spinors for a fermion with four-momentum $p_{p,\bar{p},\Lambda,\bar{\Lambda}} = (E_{p,\bar{p},\Lambda,\bar{\Lambda}},\boldsymbol{p}_{p,\bar{p},\Lambda,\bar{\Lambda}})$ and mass $m_{p,\Lambda}$ are given by}
\be
{u_p(p_p, s_p)} &=& {\frac{\Slash{p}_p + m_p}{\sqrt{2(E_p+m_p)}}(1+\gamma^5\Slash{s}_p)u_0,} \\
{v_p(p_{\bar{p}}, s_{\bar{p}})} &=& {\frac{\Slash{p}_{\bar{p}} - m_{\bar{p}}}{\sqrt{2(E_{\bar{p}}+m_{\bar{p}})}}(1+\gamma^5\Slash{s}_{\bar{p}})v_0,} \\
{u_{\Lambda}(p_{\Lambda}, s_{\Lambda}) } &=&  { \frac{\Slash{p}_{\Lambda} + m_{\Lambda}}{\sqrt{2(E_{\Lambda}+m_{\Lambda})}}(1+\gamma^5\Slash{s}_{\Lambda})u_0,} \\
{v_{\Lambda}(p_{\bar{{\Lambda}}}, s_{\bar{{\Lambda}}}) } &=& { \frac{\Slash{p}_{\bar{{\Lambda}}} - m_{\bar{{\Lambda}}}}{\sqrt{2(E_{\bar{{\Lambda}}}+m_{\bar{{\Lambda}}})}}(1+\gamma^5\Slash{s}_{\bar{{\Lambda}}})v_0,}
\ee
{where}
\be
{u_0 = \begin{pmatrix}
    1 \\
    0 \\
    0 \\
    0
\end{pmatrix},  \qquad v_0 = \begin{pmatrix}
    0 \\
    0 \\
    0 \\
    1
\end{pmatrix}.}
\ee
{Furthermore, the spin four-vectors $s_p$, $s_{\bar{p}}$, $s_{\Lambda}$, and $s_{\bar{\Lambda}}$ can be explicitly written as:}
\be
{s_p = s_{\bar{p}} = s_{\Lambda} = s_{\bar{\Lambda}} = \pm n ,}
\ee
{with $n = (0,0,1,0)$. Here, $\pm$ signs denote the two possible spin orientations of the respective particles. Additionally, we adopt the Dirac representation for the $\gamma$ matrices throughout this work. }
{For a specific spin configuration $\lambda = \{ s_p, ~s_{\bar{p}}, ~s_{\Lambda}, ~s_{\bar{\Lambda}} \}$, the corresponding amplitude $\mathcal{M}_{s_p, ~s_{\bar{p}}, ~s_{\Lambda}, ~s_{\bar{\Lambda}}}(\theta)$ is obtained by contracting the interaction operators with the Dirac spinors $u(p, \lambda), ~v(p, \lambda)$. Finally, the spin polarization $P_y(\theta)$ of outgoing $\Lambda$ hyperon can be derived from the interference between different spin states. Specifically, it is calculated by the asymmetry between the spin-up $s_{\Lambda}=+n$ and spin-down $s_{\Lambda}=-n$ production rates:}
\be 
    { P_y(\theta) \! = \! \frac{\sum\limits_{s_p,s_{\bar{p}},s_{\bar{\Lambda}}} \Big( |\mathcal{M}_{s_p,s_{\bar{p}}, +n, s_{\bar{\Lambda}}}|^2 - |\mathcal{M}_{s_p,s_{\bar{p}},-n, s_{\bar{\Lambda}}}|^2 \Big) }{\sum\limits_{s_p,s_{\bar{p}},s_{\bar{\Lambda}}} \Big( |\mathcal{M}_{s_p,s_{\bar{p}}, +n, s_{\bar{\Lambda}}}|^2 + |\mathcal{M}_{s_p,s_{\bar{p}}, -n, s_{\bar{\Lambda}}}|^2 \Big) } . } \ \ \ \ \ \ \ 
\ee 

To date, the theoretical understanding of the $p \bar{p} \to \Lambda \bar{\Lambda}$ reaction remains incomplete, as existing models have demonstrated limited capability in simultaneously describing the cross-section behavior both near the threshold and in the high-energy regions. Previous theoretical investigations have typically been constrained to reproducing either the near-threshold data or the high-energy data exclusively, leaving a significant gap in our comprehensive understanding of this process. In the present study, we address this limitation by incorporating both total and differential cross-section data into a unified framework. Through a systematic global fitting procedure, we aim to establish robust constraints on the model parameters, thereby providing a more complete and consistent description of the reaction dynamics across the entire energy spectrum.


\section{Numerical results and discussions}\label{sec:results}

\begin{figure*}[htbp]
\centering
\includegraphics[scale=0.7]{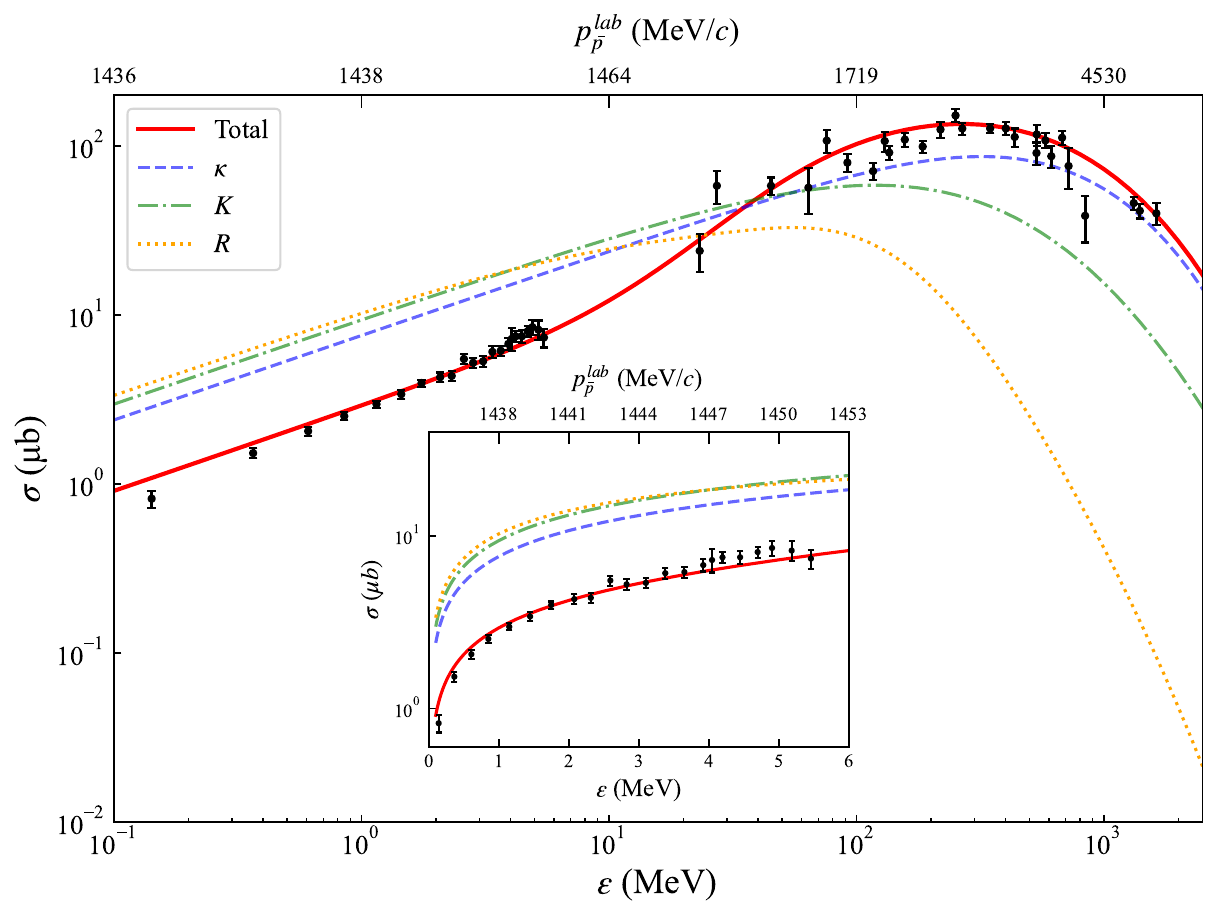}
\caption{The obtained total cross section of $p \bar{p} \to \Lambda \bar{\Lambda}$ as a function of the excess energy $\varepsilon (=\sqrt{s}-2m_{\Lambda})$, which is the excess energy of the center of mass system over the energy of the threshold $\Lambda \bar{\Lambda}$. In this figure, the blue dashed line and orange dotted line represent the contributions of $\kappa$ and $R$ resonance, respectively, while the dash-dotted curve corresponds to the $K$-exchange. The red solid line illustrates the total cross section, which is modified by the contributions of $\kappa$, $K$ and $R$ and their mutual interferences. The lower energy experimental data about 6 MeV over the $\Lambda \bar{\Lambda}$ threshold comes from Ref.~\cite{Barnes:2000be}, which can be inspected in the subfigure more clearly. The subfigure uses the same legend and axes titles as the main figure. In addition, the remaining experimental data in the higher energy region are taken from Refs.~\cite{Badier:1967zz,Oh:1973ny,Jayet:1978yq,Musgrave:1965zz,Jacobs:1977fq,Barnes:1987aw,Barnes:1989je,Barnes:1990bs,Barnes:1994bh}. Note that the antiproton laboratory momentum scale ($p^{lab}_{\bar{p}}$) is also shown in the upper sides of both figure and subfigure.}
\label{fig:cross_section}
\end{figure*}

\begin{figure*}[htbp]
\centering
\includegraphics[scale=0.35]{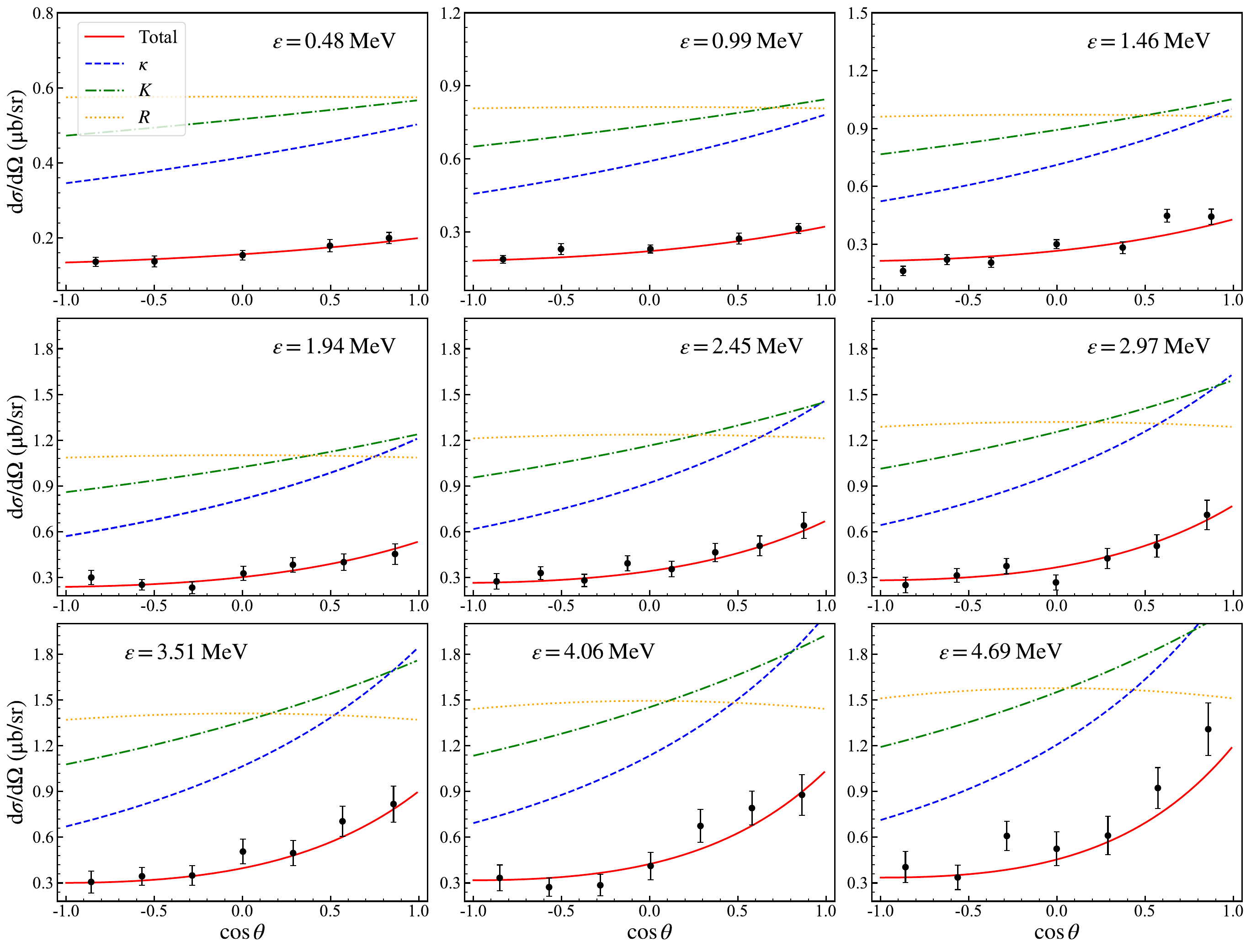}
\caption{{The obtained differential cross sections of the $p \bar{p} \to \Lambda \bar{\Lambda}$ reaction. These figures use the same legend as FIG~\ref{fig:cross_section}, and the experimental data are taken from Ref.~\cite{Barnes:2000be}.}}
\label{fig:differential_cross_section.pdf}
\end{figure*}

In this section, we present the fitted numerical results for the total and differential cross sections, and $\Lambda$ polarization for the $p \bar{p} \to \Lambda \bar{\Lambda}$ reaction. There are a total of 159 data points below excess energy $\varepsilon \simeq 1$ GeV~\footnote{The excess energy $\varepsilon$ is defined as: $\varepsilon = \sqrt{s} - 2m_\Lambda$.}. By including the current experimental data into our fitting algorithm, the fitted results are shown in TABLE~\ref{table:fit_arameters} along with Set I: $\chi^2/{\rm d.o.f} = 2.2$; Set II: $\chi^2/{\rm d.o.f} = 2.7$; Set III: $\chi^2/{\rm d.o.f} = 2.2$. There are 14, 16 and 17 free model parameters in Set I, II and III, respectively. It is found that the Set I provides the more reasonable fitting results, which leads to a mass $m_R$ below the $\Lambda\bar{\Lambda}$ threshold of 2231.4~MeV. Consequently the particle $R$ is a resonance for the $p\bar{p}$ channel but a bound state in the $\Lambda\bar{\Lambda}$ one. As a comparison, as discussed in the previous section, Set II fails to reproduce the experimental data satisfactorily, even with the including $K^*$ exchange and the additional freedom of complex couplings. This further demonstrates that the $K^*$ exchange plays a marginal role in this reaction. For the scalar $\kappa$ meson, it is turned off in Set III. Although this fit also yields a small $\chi^2/{\rm d.o.f}$ as for Set I, it results in an unphysically large width of over 1.5 GeV for the vector resonance $R$, accompanied by a strong coupling to the $\bar{p} p$ channel. Because such an extremely wide width of $R$ can not be interpreted as a genuine hadronic resonance, it indirectly illustrates $\kappa$ importance. Therefore, we may tentatively conclude that, within the current experimental data and theoretical framework, the contribution of the $K$ exchange is significantly more favorable than that of the $K^*$, while the contributions of $\kappa$ and $R$ remain indispensable. Hence, in the following we will show only the numerical results obtained with these fitted model parameters of Set I.

Before presenting the numerical results for fitted parameters of Set I, we address a potential concern. The broad decay width of the $\kappa$ resonance, $\Gamma_{\kappa} = 463 \pm 27$ MeV, should be included for the $t$-channel $\kappa$ exchange. We therefore performed a best fit including the finite widths of both the $\kappa$ and also $K^*$. However, we have found tiny changes of the fitted results. Thus, the inclusion of the widths of $t$-channel $\kappa$ and $K^*$ exchange gives negligible influence on describing the total and differential cross-section measurements for the $p\bar{p} \to \Lambda \bar{\Lambda}$ reaction over a broad energy range.

With all the model parameters determined in Set I, we can get our theoretical results. The obtained total cross section as a function of excess energy $\varepsilon$ is shown in FIG.~\ref{fig:cross_section}, from which one can easily find that the red solid line representing the total cross section successfully captures the observed experimental trend incorporating two energy regions. As for the threshold energy region, $K$ meson and $R$ resonance play significant roles, while the contribution from $\kappa$ is also important. The interference helps to modify the combined contributions from $t$-channel $K$ and $\kappa$ meson exchange and $s$-channel intermediate $R$ resonance, leading to a better reproduction of the experimental data. With the help of the higher precision in this region, there is indeed no clear evidence for the existence of a near-threshold state~\cite{Barnes:2000be}. The clearer line shape can be referred to in the subfigure in FIG.~\ref{fig:cross_section}. This is consistent with the previous studies~\cite{Barnes:2000be,Li:2021lvs}. Moreover, a preliminary conclusion regarding the near-threshold state could be confirmed if the energy gap between the two energy regions were to be filled in the future. For $\varepsilon$ below 100~MeV, contributions from $t$-channel $K$ and $\kappa$ and $s$-channel $R$ are comparable but differ at higher energy. Additionally, the resonance $R$ and the interference between $t$-channel $K$ and $\kappa$ and $s$-channel contributions optimize the peak shape together. It is expected that future experiments can be used to clarify this issue.

For the $R$ resonance, as discussed before, there have been some studies discussing the excited states of light-flavor vector mesons, particularly in terms of the resonance masses around 2200~MeV. In Ref.~\cite{Bugg:2004rj}, a partial wave analysis of $p \bar{p} \to \Lambda \bar{\Lambda}$ reaction was done, and provide evidence for a new isospin $I=0$, and spin-parity $J^{PC}=1^{--}$ resonance with mass $M_R = 2290 \pm 20$ MeV and width $\Gamma_R = 275 \pm 35$ MeV, which couples to $p\bar{p}$ and $\Lambda \bar{\Lambda}$ channels in both $S$-wave and $D$-wave. In Ref.~\cite{Wang:2021gle} the contributions of vector mesons in $e^+e^-$ annihilation to open-strangeness channels are studied, and it was reported a $\omega(3D)$ state with mass and width around 2283~MeV and 94~MeV, respectively. Ref.~\cite{Zhou:2022wwk} investigated the roles of the $\omega(3D)$ and $\omega(4S)$ states in the $e^+e^- \to \omega \eta$ and $\omega \pi^0 \pi^0$ process, and found the masses of the $\omega(3D)$ and $\omega(4S)$ states are around 2200~MeV. In addition, Refs.~\cite{Pang:2019ttv,Pang:2019ovr} also studied excited $\omega$ and $\phi$ mesons with masses close to 2200~MeV. Thus, the vector state $R$ included here with mass $2130 \pm 5$ MeV and width $292 \pm 18$ MeV, could be identified as the $\omega(3D)$ or $\omega(4S)$ state.

In addition to its good performance across the entire energy range of the total cross section, the $\kappa$ meson plays an indispensable role in the differential cross section. The results are shown in FIG.~\ref{fig:differential_cross_section.pdf}, and $\theta_{\bar{\Lambda}^*}$ is defined as the scattering angle between the three momenta of $\bar{p}$ and $\bar{\Lambda}$ in the center-of-mass frame. The center-of-mass energies are selected near the $p \bar{p}$ threshold. As the center-of-mass energy increases, the differential cross section spans a wider range of values. It's obvious to find that our results, namely the red solid curves, are in good agreement with the experimental data, {which is mainly due to the included mesons and the interference among their amplitudes.}

\begin{figure*}[htbp]
\centering
\includegraphics[scale=0.45]{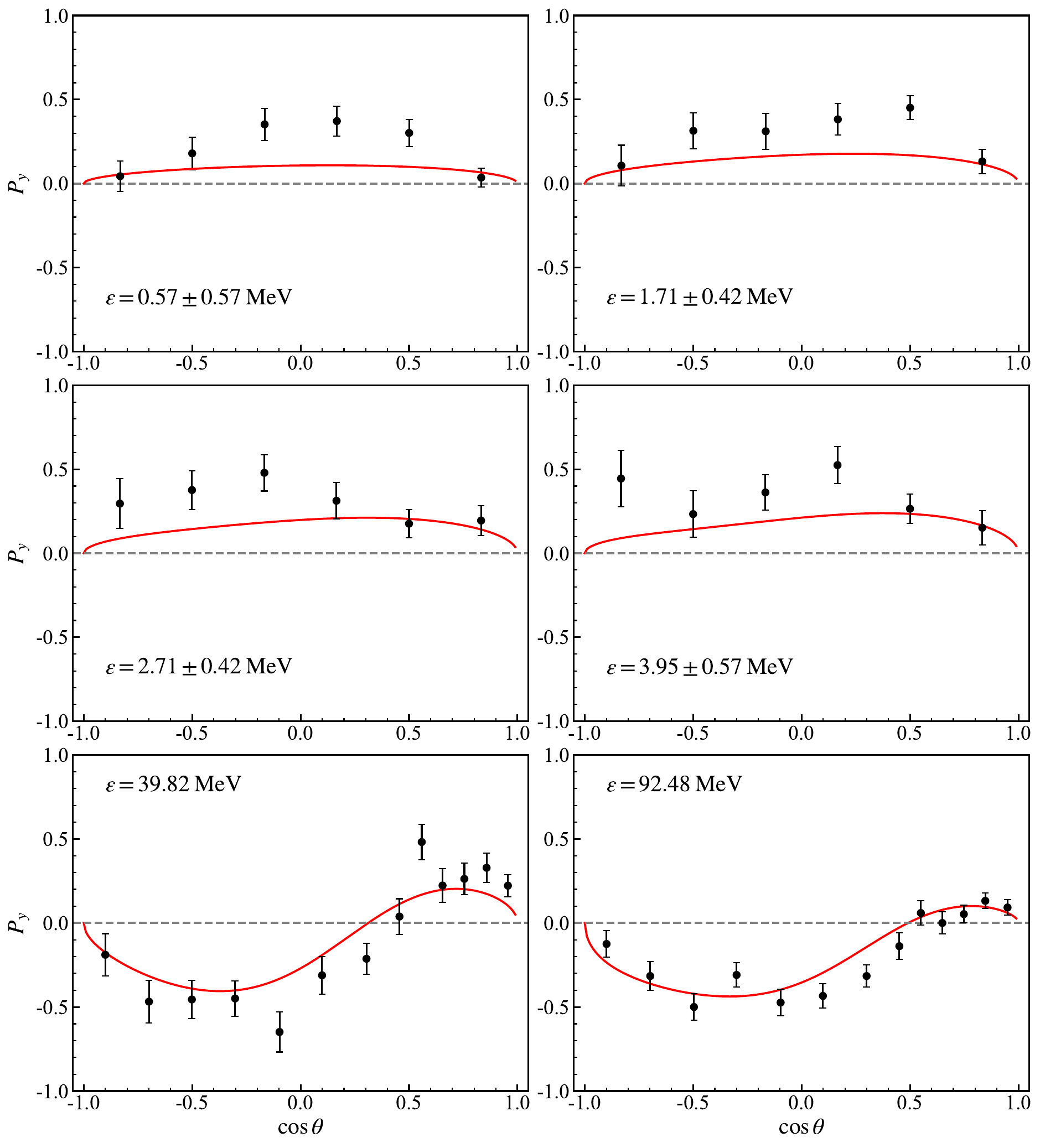}
\caption{{Polarization results of $p \bar{p} \to \Lambda \bar{\Lambda}$ reaction. The experimental data for $\varepsilon < 4$ MeV are taken from Ref.~\cite{Barnes:2000be}, whereas these data for $\varepsilon > 4$ MeV are taken from Ref.~\cite{Barnes:1990bs}.}}
\label{fig:polarization.pdf}
\end{figure*}

The calculated results for the polarization observable $P_y(\theta)$ are compared with the experimental data in FIG.~\ref{fig:polarization.pdf}. The comparison covers a wide kinematic range from the near-threshold region to higher energies. At low excess energies ($\varepsilon < 4$ MeV), the measured polarization is consistently positive across the entire angular range, showing a gentle angular dependence. Our model qualitatively reproduces this feature, although the magnitude is slightly underestimated at intermediate angles for $\varepsilon = 1.71$ and $2.71$ MeV. As the energy increases to $\varepsilon = 39.82$ and $92.48$ MeV, the polarization exhibits a distinct structure characterized by a sign change. $P_y(\theta)$ becomes negative in the $\cos\theta < 0.5$ backward direction and $\cos\theta > 0.5$ positive in the forward direction. Our theoretical calculation captures this characteristic angular distribution and the zero-crossing behavior remarkably well, indicating that the interference between the contributing amplitudes is correctly described with our theoretical mechanism. It was also found that reproducing these polarization data for the $p \bar{p} \to \Lambda \bar{\Lambda}$ reaction requires a large interference between $t$-channel $\kappa$ and $K$ exchange, as well as the $s$-channel intermediate resonance $R$.

As we mentioned earlier, previous studies on the $t$-channel reaction $p \bar{p} \to \Lambda \bar{\Lambda}$ have mainly focused on meson exchanges involving $K$ and $K^*$~\cite{Carbonell:1993dt,Shyam:2014dia, Mueller-Groeling:1990uxr, Haidenbauer:1991kt, Haidenbauer:1992wp}. However, those models are limited to specific energy regions and do not provide a comprehensive description across the entire energy spectrum relevant to current experimental data. Actually, we have also considered the contributions of $K$ or $K^*$ mesons based on the effective Lagrangian methods. However, as anticipated, neither of the two mesons can capture the total cross sections near the threshold and high-energy behavior simultaneously. Beyond achieving a consistent total cross section across both low and high energy regions, it is essential for the model to reproduce the corresponding differential cross section data, as this gives a critical criterion for checking its validity. We found that $K$ and $K^*$ failed to reproduce the differential cross section data even if they could give a rough description of the total cross section near the reaction threshold. {The introduction of the $\kappa$ meson has provided a promising way, leading us to exclude $K^*$ rather $K$ mesons from our model. Besides, we have also studied contributions from the narrow state $X(2231)$ proposed in Ref.~\cite{Li:2021lvs} that has strong coupling to the $\Lambda \bar{\Lambda}$ channel, it was found that its contribution may not be important because of the limited experimental data, whereas another resonance $R$ with a higher mass proposed in this work appears to play a more essential role within our model.} Nevertheless, more precise experimental data will test our model calculations. Furthermore, those new experimental data will be used as essential inputs to improve the theoretical descriptions of the two-body baryon–antibaryon interactions.


\section{Summary} \label{sec:summary}

{In this study, we conduct a comprehensive theoretical analysis of the reaction $p \bar{p} \to \Lambda \bar{\Lambda}$ by combining the total and differential cross-section data near the threshold and in the high-energy regions, as well as the produced $\Lambda$ polarization data. Employing the framework of the effective Lagrangian approach, we incorporate the $t$-channel exchange of the $\kappa$ and $K$ mesons, together with the $s$-channel contributions from a vector $R$ resonance.} Our model successfully reproduces the experimental data, offering a robust description of the observed phenomena. This result highlights a distinct difference in the production mechanism and the final state interaction of the $p \bar{p} \to \Lambda \bar{\Lambda}$ reaction, providing new insights into the underlying dynamics of this process.

Compared to the general consideration of $t$-channel exchanges of $K$ and $K^*$ mesons, the inclusion of $\kappa$ meson proves more effective in simultaneously describing the total cross sections both in the threshold and high-energy regions. Moreover, the $\kappa$ exchange shows clear advantages in reproducing the differential cross sections. In addition, our theoretical calculations also provide a satisfactory description of the available hyperon polarization data, indicating that the underlying reaction mechanism is consistently captured across multiple observables. Therefore, we consider $\kappa$ an indispensable component in the theoretical study of the $p \bar{p} \to \Lambda \bar{\Lambda}$ reaction. The total cross section experimental data provide hints of an excited vector state, which provides significance in the overall fitting to the experimental results. Our conventional model, with leading $t$-channel $\kappa$ and $K$ exchange supplemented by a s-channel $R$ resonance, is simpler than the complicated quark model approach of Ref.~\cite{Ortega:2011zza}. And it is anticipated that our findings will offer new perspectives and insights for experiments aimed at refining the understanding of the $p \bar{p} \to \Lambda \bar{\Lambda}$ reaction, thereby enhancing our knowledge of hyperon production and the dynamics of light hadrons in the low-energy region.

More precise experimental measurements on the $p \bar{p} \to \Lambda \bar{\Lambda}$ reaction, especially for the excess energy about $10$ to $100$ MeV, are expected to provide valuable insights into its reaction kinematics. These experimental measurements can be carried out by the next-generation facilities~\cite{Yang:2013yeb,ZHAO:2020llg,Li:2025fnp}.

\section*{ACKNOWLEDGEMENT}

This work is partly supported by the National Key R\&D Program of China under Grant No. 2023YFA1606703 and the National Natural Science Foundation of China under Grant Nos. 12435007, 12361141819, 12475086, 12192263, and 12575094.

\normalem
\bibliography{ref}

\end{document}